\documentclass[a4paper]{article}
  
\usepackage{INTERSPEECH2021}
\title{Classification of Autistic and Non-Autistic Children’s Speech: A Cross-Linguistic Study in Finnish, French, and Slovak}
\name{Sofoklis Kakouros$^1$, Ida-Lotta Myllylä$^2$}

\address{
  $^1$Speech and Cognition Research Group, Signal Processing Research Centre, Tampere University\\
  $^2$Department of Languages, University of Helsinki}
\email{sofoklis.kakouros@tuni.fi,ida-lotta.myllyla@helsinki.fi}

\begin{document}

\maketitle
\begin{abstract}
We present a cross-linguistic study of speech in autistic and non-autistic children speaking Finnish, French, and Slovak. We combine supervised classification with within-language and cross-corpus transfer experiments to evaluate classification performance within and across languages and to probe which acoustic cues are language-specific versus language-general. Using a large set of acoustic-prosodic features, we implement speaker-level classification benchmarks as an analytical tool rather than to seek state-of-the-art performance.

Within-language models, evaluated with speaker-level cross-validation, yielded heterogeneous results. The Finnish model performed best (Accuracy 0.84, F1 0.88), followed by Slovak (Accuracy 0.63, F1 0.68) and French (Accuracy 0.68, F1 0.56). We then tested cross-language generalization. A model trained on all pooled corpora reached an overall Accuracy of 0.61 and F1 0.68. Leave-one-corpus-out experiments, which test transfer to an unseen language, showed moderate success when testing on Slovak (F1 0.70) and Finnish (F1 0.78), but poor transfer to French (F1 0.42). Feature-importance analyses across languages highlighted partially shared, but not fully language-invariant, acoustic markers of autism.

These findings suggest that some autism-related speech cues generalize across typologically distinct languages, but robust cross-linguistic classifiers will likely require language-aware modeling and more homogeneous recording conditions.

\end{abstract}

\section{Introduction}

Autism as a neurodevelopmental condition is characterized by differences in social communication and interaction \cite{lord2018autism}. Pragmatic language and speech may deviate from expected conversational norms \cite{grice2023linguistic}, and the prosody of autistic speakers has been noted to differ from that of non-autistic peers \cite{fusaroli2022toward}. Speaker-related differences and heterogeneity of prosodic features observed in the autistic population are significant, but evidence points to pitch characteristics being commonly deviating features \cite{fusaroli2017voice,fusaroli2022toward}. Pitch may be especially varied or exaggerated, albeit flat pitch may also be observed, pointing to prosodic extremes as salient depictors of autistic prosody \cite{wiklund2019, fusaroli2022toward, okuizumi2025acoustic}. Other prosodic features include differences in controlling intensity levels, atypical pausing and durational prosody, and atypical voice quality \cite{trayvick2024speech, fusaroli2017voice, fusaroli2022toward}. Mounting empirical evidence points to differing prosody in autism, and autism-adjacent prosodic profiles have been identified both within various languages and cross-linguistically \cite{fusaroli2022toward,lau2022cross,asghari2021distinctive}. Using automatic classification methods, autism has been detected from speech signals using various acoustic features related to fundamental frequency, intensity, and temporal and spectral features \cite{lau2022cross, chi2022classifying,mohanta2022analysis}. 

In this study, we employed supervised models trained on prosodic features \cite{kakouros23_interspeech} to classify autistic and non-autistic speech within and across languages. By comparing model performance in monolingual and cross-linguistic setups, and by analyzing feature importance in both conditions, we identified possible autism-adjacent prosodic cues that may be language-dependent as well as cues that may be shared across languages.


\section{Materials}
We used three datasets in our experiments, one for each language under study: Finnish, French, and Slovak. Descriptions of the data and preprocessing follow in the next subsections (see also Table \ref{tab:data_summary} for a summary of the data).

\subsection{Finnish data}
Finnish speech data were collected from six native Finnish-speaking autistic males (aged 11–13) who were diagnosed with Asperger’s syndrome (ICD-10 \cite{ICD10_2004}), a diagnosis since folded under the diagnosis of ASD. The control (TD) data were collected from six age-matched, non-autistic, native Finnish-speaking males with no reported developmental concerns. 

The autistic participants were recorded during spontaneous group discussions in a hospital-based social skills and communication intervention in the Helsinki metropolitan area in Finland. Three participants attended one session, and three another. Each session lasted approximately two hours, comprising free-form discussion and games. The sessions were led by two healthcare professionals.

The control group was recorded in a comparable teacher-led discussion at a Helsinki metropolitan area school. This conversational setting was constructed for the purposes of research, and it consisted of free-form discussion. 

Participants for both groups were selected based on availability from their respective locations. All sessions were recorded in quiet environments using portable recorders and head-mounted microphones for each speaker, capturing 16-bit mono audio at 44.1 kHz. 

\subsection{French data}
French speech data were collected from six native French-speaking autistic males and three non-autistic controls, all aged 11–13. The autistic participants were recorded during spontaneous group sessions in a speech therapy clinic, an intervention for social skills and communication. Two participants attended one session, and four another. Each session lasted approximately an hour, again comprising free-form discussion and games. Sessions were led by two speech therapists. The session with three speakers was recorded in Geneva, Switzerland and the other in Orbe, Switzerland. 

The control group was recorded in an adult-led discussion in Lausanne, Switzerland, a conversational setting constructed for the purposes of research, which again comprised free-form discussion. Participants for both groups were selected based on availability from their respective locations. All sessions were recorded in quiet environments using portable recorders and head-mounted microphones for each speaker, capturing 16-bit mono audio at 44.1 kHz. 

\subsection{Slovak data}
Slovak speech data utilized in this study is the Slovak Autistic and Non-Autistic Child Speech Corpus (SANACS) \cite{kruyt2024slovak}. It consists of recordings of collaborative, task-oriented conversations between children (autistic or non-autistic) and a non-autistic adult experimenter, who are all native speakers. The data entail speech from 67 Slovak children aged 6-12 (mean 9.2). Of the speakers, 37 were autistic (7 girls, 30 boys) and 30 non-autistic (7 girls, 23 boys). Each child interacted with the same experimenter. All autistic children were diagnosed with ASD (autism spectrum disorder). The Maps task was used elicit this corpus. Most tasks consisted of six trials: a practice and two trials where the experimenter is the describer and the child the follower, followed by one practice and two trials in which the roles were reversed. All sessions were recorded with portable recorders and portable microphones capturing stereo audio, where channel 1 is the experimenter and channel 2 the child. The corpus was recorded in collaboration between the Institute of Informatics, Slovak Academy of Sciences, and the Academic Research Center for Autism, Comenius University, both in Bratislava, Slovakia.


\subsection{Data preprocessing}
The speech data of all adult session leaders were not included in the analysis. For the Slovak data, the child speech was extracted to mono from the stereo audio. All speech data from the child speakers were segmented into IPUs (interpausal units), meaning sections of speech between pauses of at least 200 ms, produced by one speaker. The segmenting was done automatically using intensity thresholds and validated manually. IPUs with overlapping speech and those consisting only of non-lexical vocalizations, were discarded prior to the analysis. One speaker from the Slovak dataset was excluded at this stage due to a substantial amount of overlapping speech and IPUs consisting only of non-lexical hesitation markers.

\section{Methods}

\subsection{Feature extraction}
All speech recordings were processed with the openSMILE toolkit, an open-source toolkit for audio feature extraction, using the eGeMAPS utterance-level functionals configuration \cite{eyben2015geneva}, producing an 88-dimensional feature vector for each inter-pausal unit (IPU). This configuration combines selected low-level descriptors with their statistical functionals, covering fundamental frequency ($f0$), intensity, spectral characteristics, and voice quality measures (e.g., jitter, shimmer). The extracted features provide a comprehensive representation of prosodic dynamics and voice quality characteristics capturing pitch and energy distribution, spectral balance, and temporal variability.

Feature outputs were compiled into a single dataset, which also included metadata for each IPU (e.g., speaker ID, group, language, sex and age). These metadata fields were stored separately from the 88-dimensional acoustic–prosodic feature vector and associated with each IPU on the basis of their filename.

\begin{table}[t]
\centering
\small
\setlength{\tabcolsep}{3pt}
\resizebox{\columnwidth}{!}{%
  \begin{tabular}{lcccc}
    \toprule
    Language & ASD (utt./IDs) & TD (utt./IDs) & Mean dur. (s) & Total (utt./IDs) \\
    \midrule
    FINNISH & 1323 / 6  & 249 / 6   & 2.11 & 1572 / 12 \\
    FRENCH  & 884 / 6   & 659 / 3   & 1.79 & 1543 / 9  \\
    SLOVAK  & 2867 / 37 & 2230 / 29 & 2.36 & 5097 / 66 \\
    \midrule
    ALL     & 5074 / 49 & 3138 / 38 & 2.20 & 8212 / 87 \\
    \bottomrule
  \end{tabular}
}
\caption{Summary of utterances (utt.), unique speaker IDs per diagnostic group, and mean recording duration by language.}
\label{tab:data_summary}
\vspace{-3.0\baselineskip}
\end{table}

\subsection{Classification benchmarks}
To better understand both cross-linguistic differences and distinctions between autisic and non-autistic speech, we implemented a straightforward supervised classification framework to assess within-language and across-language generalization. Specifically, we trained simple classifiers on the 88-dimensional acoustic feature set extracted with openSMILE, using XGBoost \cite{chen2016xgboost} and Random Forest models to discriminate between autistic and non-autistic speakers' samples. Within-language experiments evaluated how well models trained and tested on the same language could capture language-specific acoustic markers, while across-language experiments examined the extent to which models trained on one language (or a subset of languages) generalized to another, thereby probing the robustness and transferability of the identified cues across linguistic contexts. 

It is important to emphasize that our objective is not to establish state-of-the-art performance for ASD vs. TD classification in these languages. Instead, we deliberately focus on simple and interpretable classifiers applied to a set of well-established acoustic-prosodic features. This choice allows us to prioritize transparency and explanatory power over raw predictive accuracy, using these models as analytical tools to gain a clearer understanding of how specific prosodic and acoustic characteristics relate to diagnostic group differences across languages.

\subsection{Feature importance}
For the feature importance analysis, we employed a diverse set of model-based and model-agnostic approaches to obtain robust and interpretable estimates. Specifically, we relied on tree-based ensemble methods, including XGBoost, Random Forest, and single Decision Trees, to derive importance scores based on split gains and impurity reductions. To complement these model-specific metrics and provide more theoretically grounded attributions, we applied TreeSHAP \cite{lundberg2020local} to quantify the marginal contribution of each feature to the model output, and we further validated these findings using permutation-based feature importance, which assesses the impact of randomly shuffling individual features on predictive performance. All analyses were conducted both separately for each language and on the combined multilingual dataset, enabling us to examine language-specific patterns as well as globally stable predictors.

\section{Experiments}

\subsection{Classification}

\subsubsection{Within-language analysis}
For the within-language setting, we train separate models for each language (Slovak, Finnish, and French). To obtain realistic estimates of how well the models generalize to unseen individuals, we perform cross-validation at the speaker level, ensuring that both ASD and TD speakers are represented in both the training and test sets. In practice, we define folds over unique subject IDs and ensure that all utterances from a given speaker are assigned either to the training set or to the test set in a given fold, but never to both. We use a nominal five-fold setup; when the number of speakers or the class distribution is limiting, the effective number of folds is adjusted automatically to maintain valid splits while keeping at least two folds. The key point is that model evaluation always reflects performance on speakers the model has not seen during training. 

Because our dataset is substantially imbalanced with respect to gender (most recordings come from male participants), we cannot reliably enforce gender-balanced splits. Instead, the partitioning is randomized at the level of unique subject IDs only. The classifiers are trained with class weighting to partially compensate for ASD–TD label imbalance (see also Table \ref{tab:data_summary}). However, this imbalance cannot be fully corrected across all datasets, as the Finnish data are particularly skewed. For each language we then run both Random Forest and XGBoost and report standard accuracy and F1-score, averaged across folds (mean and standard deviation).

\subsubsection{Cross-linguistic analysis}
To examine cross-language generalization, we use two complementary setups: (1) pooled multilingual training and (2) leave-one-corpus-out (LOCO) evaluation. In the first setup, we pool data from all three languages into a single dataset and again perform 5-fold cross-validation at the speaker level (with the same constraint that speakers do not appear in both training and test folds). This setup asks how well a single model can exploit a larger and more diverse training set and whether exposure to multiple languages leads to a classifier that is robust across language-specific acoustic idiosyncrasies. It also provides a point of comparison with the language-specific baselines: if the pooled model performs similarly or better, this suggests that cross-linguistic information is beneficial rather than harmful.

In the second cross-language scheme, we perform a leave-one-corpus-out (LOCO) evaluation, which more directly probes cross-lingual transfer. Here we treat each language as a separate corpus. At each iteration we hold out one language entirely for testing and train the models on the remaining two languages. For example, we may train on Slovak and Finnish and test on French, then train on Slovak and French and test on Finnish, and so on until each language has served as the unseen test set once. Unlike the pooled setup, there is no mixing of languages between train and test in LOCO: the model is evaluated on a language that it has never encountered during training. This configuration addresses a different question: to what extent are the acoustic–prosodic markers that distinguish ASD from TD stable enough across languages that a model trained on some languages can generalize to a completely new one. Good performance in the LOCO setting would indicate that at least some ASD–TD characteristics captured by the eGeMAPS feature set are language-independent, which in turn suggests that models trained on better-resourced languages could be applied, with caution, to languages for which little or no data are available.

\subsection{Feature importance}
\begin{figure}[t]
    \centering
    \includegraphics[width=\columnwidth]{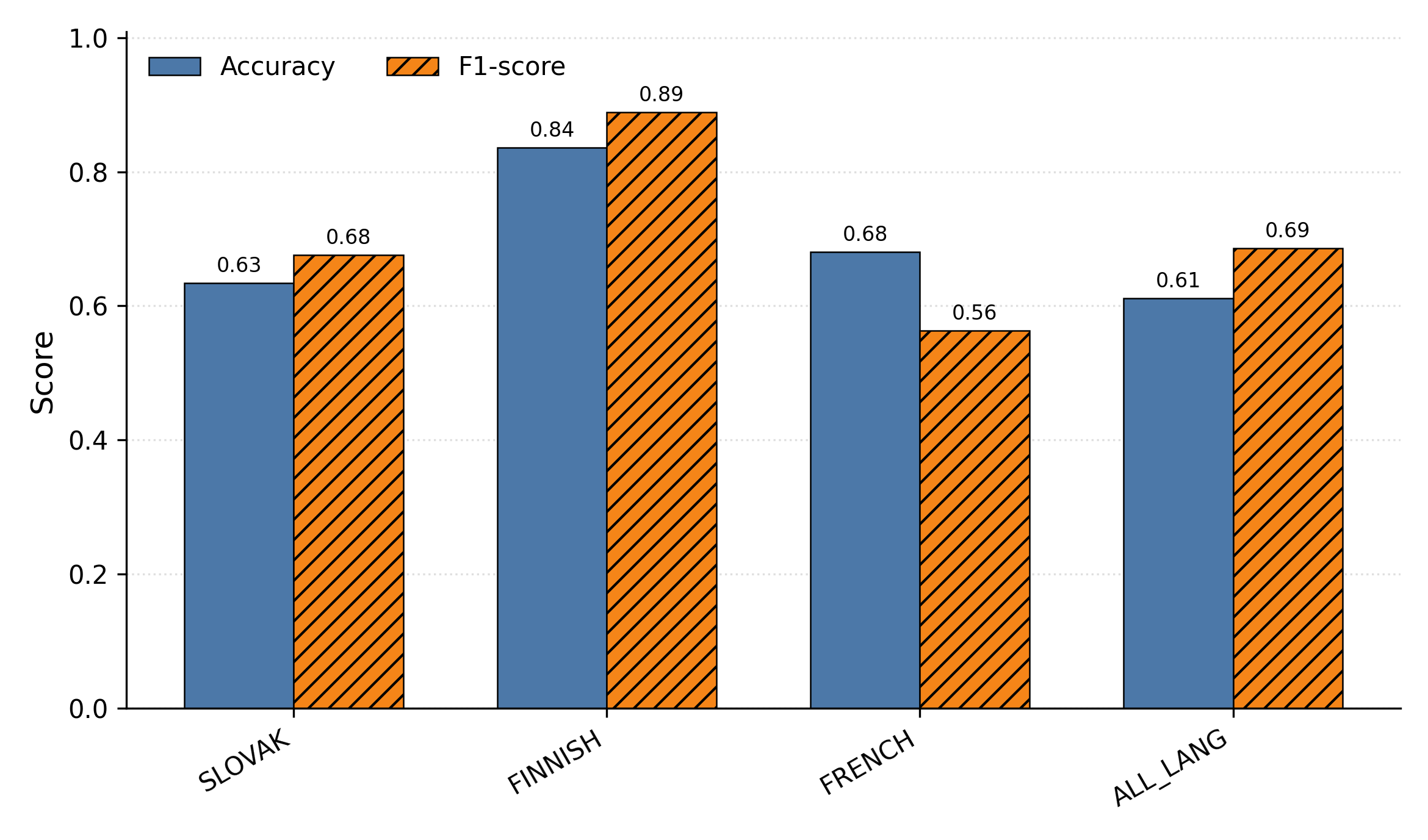}
    \caption{Accuracy and F1-score per language and overall for the XGBoost model.}
    \label{fig:accuracy_f1_per_language_all_xgboost}
\vspace{-4mm}
\end{figure}

\begin{figure}[t]
    \centering
    \includegraphics[width=\columnwidth]{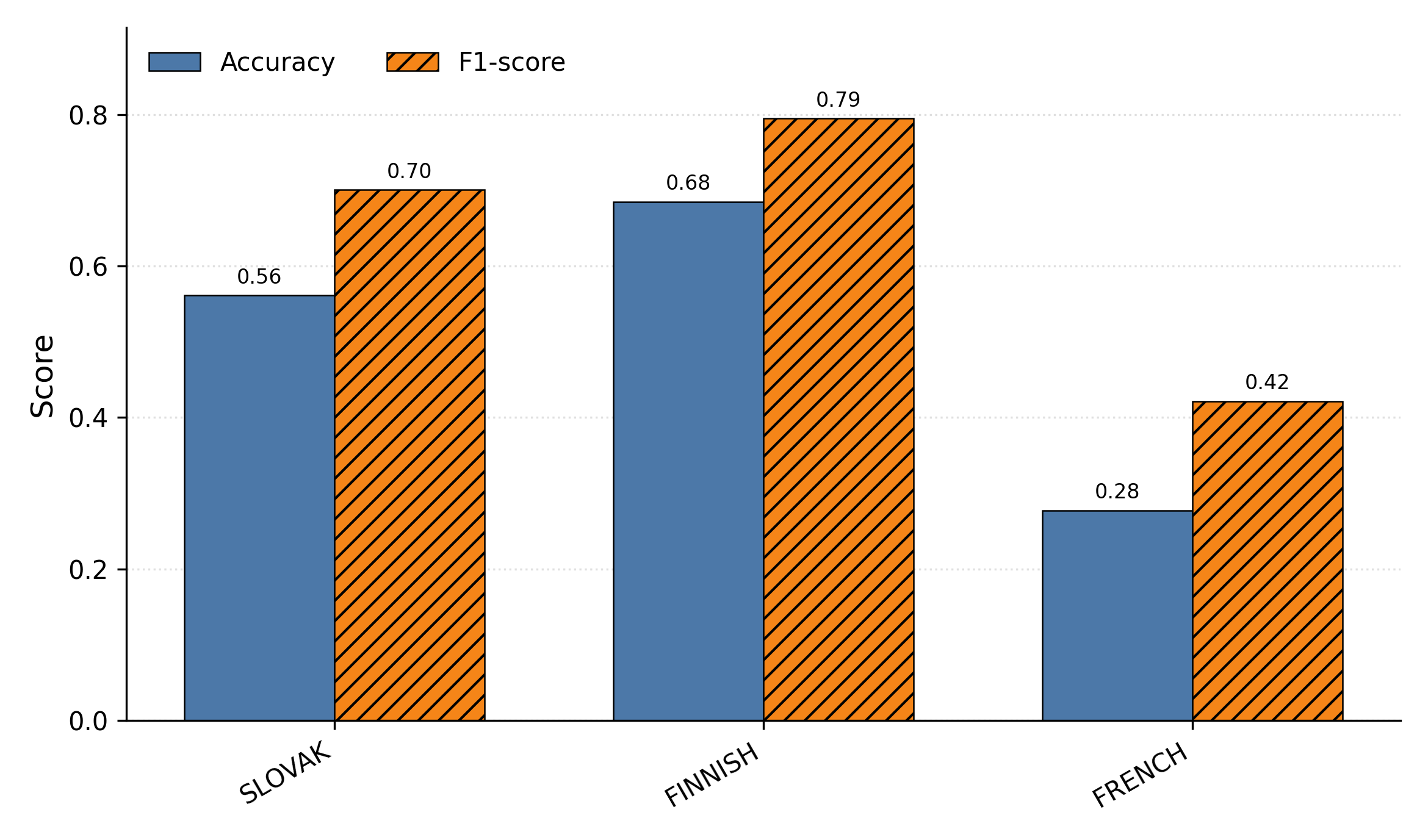}
    \caption{LOCO accuracy and F1-score per language for the XGBoost model.}
    \label{fig:accuracy_f1_loco_xgboost}
\vspace{-6mm}
\end{figure}

\subsubsection{Within-language analysis}
\begin{figure}[t]
    \centering
    \includegraphics[width=\columnwidth]{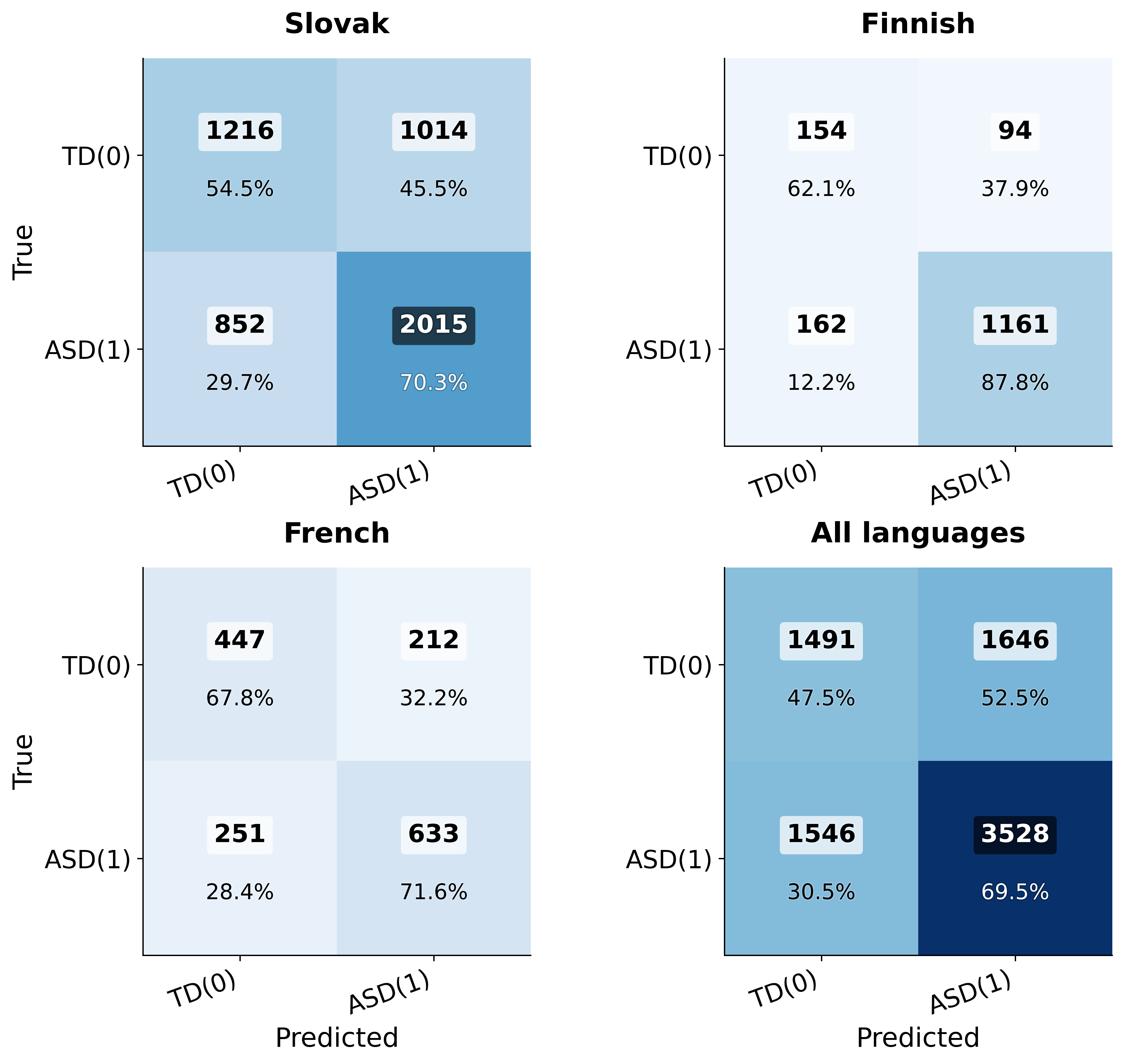}
    \caption{CM per language and overall for the XGBoost model.}
    \label{fig:cm_per_language_all_within}
\vspace{-3mm}
\end{figure}

\begin{figure}[t]
    \centering
    \includegraphics[width=\columnwidth]{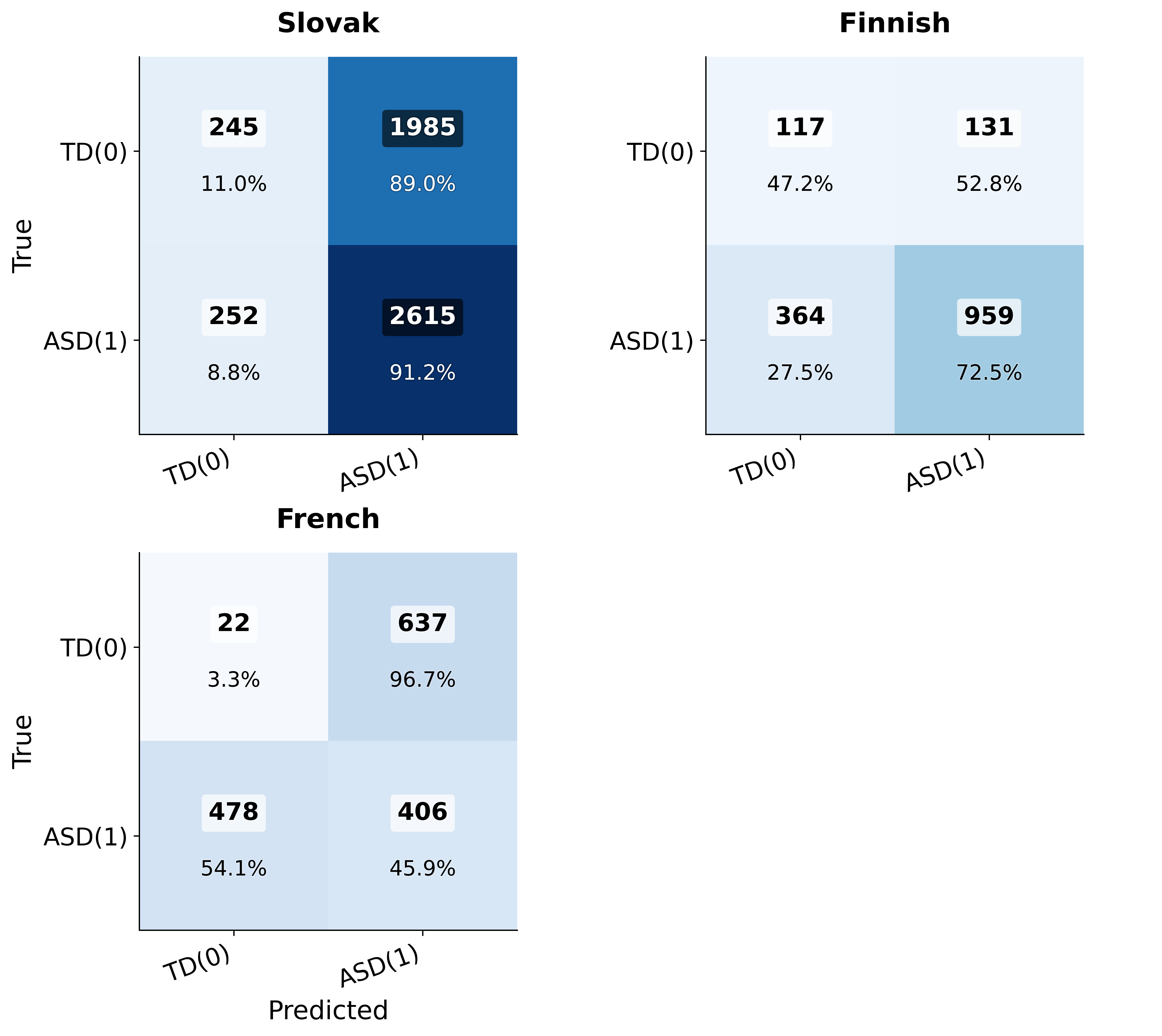}
    \caption{CM for LOCO per language for the XGBoost model.}
    \label{fig:cm_loco}
\vspace{-6mm}
\end{figure}

\begin{figure}[t]
    \centering
    \includegraphics[width=\columnwidth]{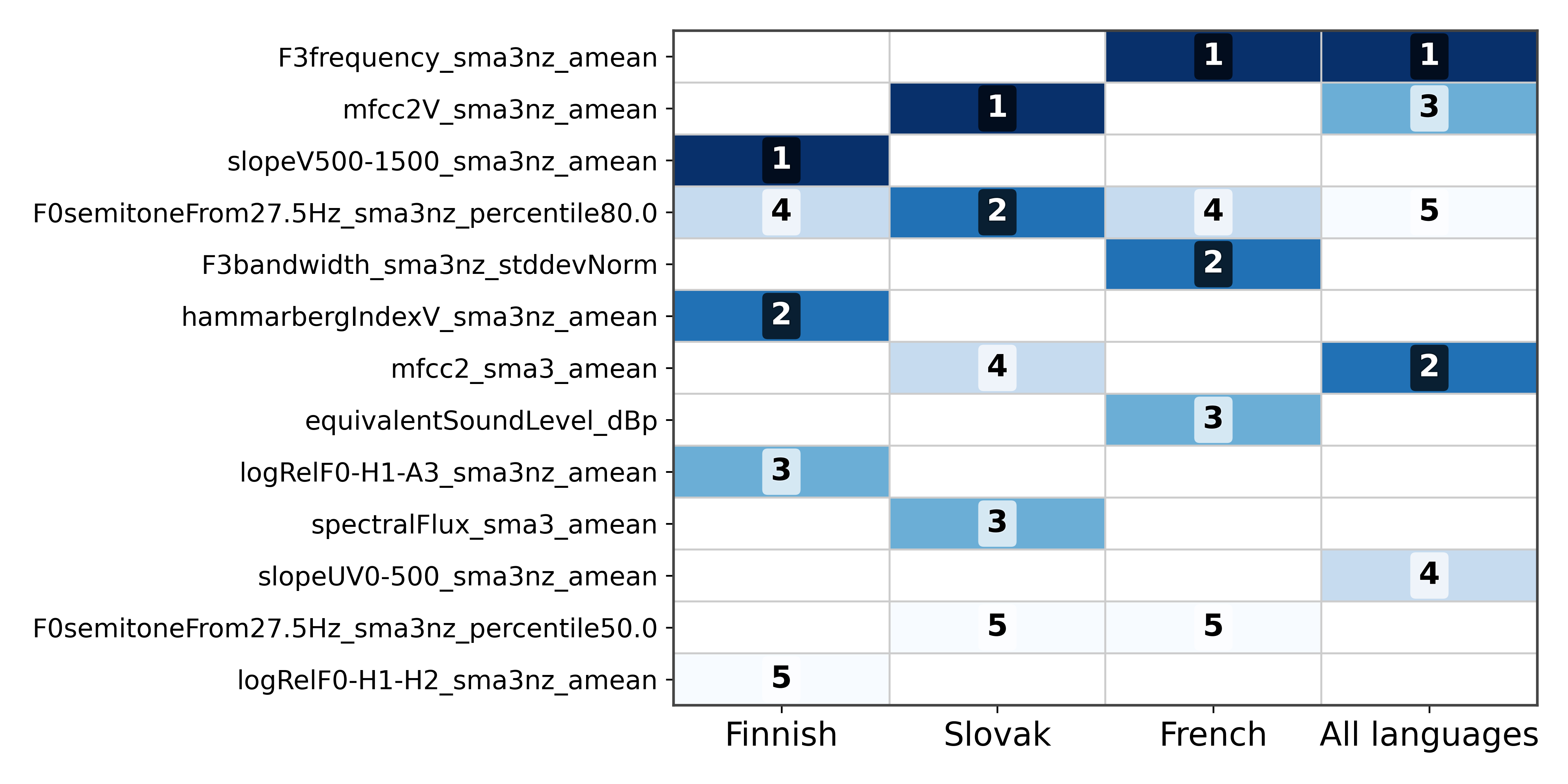}
    \caption{Consensus feature importance per language and for all languages.}
    \label{fig:consensus_features}
\vspace{-6mm}
\end{figure}

For feature importance, we adopt a consensus-based approach to better capture the most relevant features. Feature importance is estimated using decision trees, Random Forest, XGBoost, TreeSHAP, and permutation importance (see also \cite{kakouros25b_interspeech}). For each feature, we aggregate the rankings produced by these methods to obtain an average (and best) rank. We then select features that are deemed important by at least two methods, order them according to their average/best rank, and derive a consensus top-5 set of features for each scope and language.

\subsubsection{Cross-linguistic analysis}

For the cross-linguistic analysis, we employ the same procedure as in the within-language analysis, but extend its application to the full dataset comprising all three languages under investigation. This allows us to assess whether the patterns observed within individual languages also hold when the data are considered jointly across languages. 

\section{Results and Discussion}
In the next subsections, for each within- and cross-linguistic analysis, we present classification outcomes for a single model (XGBoost, since differences from RandomForest were generally small) as well as the corresponding feature importance.

\subsection{Within-language analysis}
The within-language classification results indicate that a simple classification setup can distinguish between the ASD and TD classes, although performance varies across languages as can be seen in Figure \ref{fig:accuracy_f1_per_language_all_xgboost}. The best overall performance was obtained for Finnish (84\% accuracy, 89\% F1), followed by Slovak (63\% accuracy, 68\% F1) and French (68\% accuracy, 56\% F1). The Finnish results should, however, be interpreted with caution. First, the dataset was substantially imbalanced, with the ASD class containing many more samples (see Table \ref{tab:data_summary}), which is also reflected in the confusion matrix (Figure \ref{fig:cm_per_language_all_within}). Second, we observed qualitative differences in the speaking styles of the ASD and TD groups in Finnish, possibly due to greater familiarity with the clinician in the ASD group, which appeared to elicit more dynamic or expressive language use rather than simple responses or short statements. Overall, the results for French and Slovak appear more reliable and robust. This is especially true for Slovak, where the class distributions are approximately balanced and a relatively large and diverse dataset is available.

With respect to feature importance, the results are presented in Figure \ref{fig:consensus_features}. Across Finnish, Slovak, and French, F0 distribution (pitch level and range) is consistently important for distinguishing ASD from TD speech, while the secondary cues differ by language: Finnish relies mainly on spectral tilt (see, e.g., \cite{kakouros2018comparison}) and voice quality, Slovak on global spectral shape and dynamics, and French on higher-formant structure and overall intensity. 


\subsection{Cross-linguistic analysis}

Our cross-linguistic analysis indicates only modest generalization with our simple setup, as shown in Figures~\ref{fig:accuracy_f1_per_language_all_xgboost}, ~\ref{fig:accuracy_f1_loco_xgboost},  ~\ref{fig:cm_per_language_all_within} and ~\ref{fig:cm_loco}. When pooling data from all languages, the model reaches a mean performance across splits of 61\% accuracy and 69\% F1. Under the LOCO setting, however, performance varies widely across languages: the best results are obtained for Finnish (68\% accuracy, 79\% F1), followed by Slovak (56\% accuracy, 70\% F1) and French (28\% accuracy, 42\% F1). Overall, these results suggest that the languages do share some task-relevant information but, consistent with the feature-importance analysis, they may not rely on the same features. Note also that the large differences between accuracy and F1 in the LOCO settings are due to class imbalance (see Figure ~\ref{fig:cm_loco}).

With respect to feature importance, in the case of the pooled data across all languages, the features from Figure~\ref{fig:consensus_features} suggest that both segmental (vowels and consonants) and pitch properties contribute in a relatively language-general way. Overall, language-specific analyses reveal somewhat different emphases, but the pooled model highlights a shared, language-general set of prosodic and spectral–segmental cues that consistently distinguish ASD from typical speech (features related to pitch and spectral shape).



\section{Conclusions}

Overall, our findings show that supervised models trained on prosodic features can distinguish ASD from TD speech within languages reasonably well, and to a more limited extent across languages, supporting the presence of both language-specific and language-general autism-adjacent prosodic cues. Cross-linguistically, pooled and LOCO analyses reveal only modest generalization and substantial variation by language, suggesting that while some task-relevant information is shared, it is not encoded in fully overlapping acoustic–prosodic patterns. Consistent with this, feature-importance analyses identify F0 distribution as a robust language-general cue, alongside language-specific  spectral, voice-quality, and segmental features, indicating that prosodic markers of ASD speech are partly general but also shaped by language-specific characteristics.

\section{Acknowledgements}

SK was supported by the L-SCALE project funded by Kone Foundation. ILM was funded by the Research Council of Finland under grant number 362906. The authors wish to acknowledge CSC – IT Center for Science, Finland, for providing the computational resources.

\bibliographystyle{IEEEtran}

\bibliography{mybib}


\end{document}